\documentclass[11pt,floatfix,showpacs]{revtex4}
\usepackage[dvips]{graphicx,color}
\usepackage{amsmath}
\usepackage{amsfonts}
\usepackage[latin1]{inputenc}

\begin{document}

\title{Deconfinement and chiral restoration in nonlocal
PNJL models at zero and imaginary chemical potential}

\author{V. Pagura$^{a,b}$, D. G\'omez Dumm$^{b,c}$ and N.N.Scoccola$^{a,b,d}$}

\address{$^{a}$ Physics Department, Comisi\'on Nacional de
Energ\'{\i}a Atómica, Av.Libertador 8250, 1429 Buenos Aires,
Argentina \\
$^{b}$ CONICET, Rivadavia 1917, 1033 Buenos Aires, Argentina \\
$^{c}$ IFLP, CONICET $-$ Dpto.\ de F\'{\i}sica, Universidad Nacional
de La Plata, C.C. 67, 1900 La Plata, Argentina,\\
$^{d}$ Universidad Favaloro, Sol{\'\i}s 453, 1078 Buenos Aires,
Argentina}

\begin{abstract}
We study the deconfinement and chiral restoration transitions in the context
of non-local PNJL models, considering the impact of the presence of
dynamical quarks on the scale parameter appearing in the Polyakov potential.
We show that the corresponding critical temperatures are naturally entangled
for both zero and imaginary chemical potential, in good agreement with
lattice QCD results. We also analyze the Roberge Weiss transition,
which is found to be first order at the associated endpoint.
\end{abstract}


\maketitle

The detailed understanding of the behavior of strongly interacting matter at
finite temperature and baryon density represents an issue of great interest
in particle physics~\cite{Rischke:2003mt}. From the theoretical point of
view, this problem can be addressed through lattice QCD
calculations~\cite{All03,Fod04,Kar03}, which have been significantly
improved in the last years. However, this ab initio approach is not yet able
to provide a full understanding of the QCD phase diagram. One well-known
difficulty is given by the so-called sign problem, which arises when dealing
with finite real chemical potentials. Thus, it is worth to develop
alternative approaches, such as the study of effective models that show
consistency with lattice QCD results and can be extrapolated into regions
not accessible by lattice techniques. One of these effective theories,
proposed quite recently, is the so-called Polyakov-Nambu-Jona-Lasinio (PNJL)
model~\cite{Meisinger:1995ih,Fukushima:2003fw,Megias:2004hj,Ratti:2005jh,Roessner:2006,Mukherjee:2006hq,Sasaki:2006ww},
an extension of the well-known NJL model~\cite{reports} in which quarks are
coupled to the Polyakov loop (PL), providing a common framework to study
both the chiral and deconfinement transitions. As a further improvement over
the (local) PNJL model, extensions that include covariant non-local quark
interactions have also been
considered~\cite{Blaschke:2007np,Contrera:2007wu,Hell:2008cc}. The non-local
character of the interactions arises naturally in the context of several
successful approaches to low-energy quark dynamics, and leads to a momentum
dependence in the quark propagator that can be made
consistent~\cite{Noguera:2008} with lattice results. It has been
shown~\cite{BB95,BGR02,Scarpettini:2003fj,GomezDumm:2006vz} that non-local
models provide a satisfactory description of hadron properties at zero
temperature and density. Moreover, it has been found that, under certain
conditions, it is possible to derive the main features of non-local PNJL
models starting directly from QCD~\cite{Kondo:2010ts}. Related
Polyakov-Dyson-Schwinger equation models have also been recently
considered~\cite{Horvatic:2010md}.

The aim of the present work is to analyze the relation between the
deconfinement and chiral restoration transitions at both zero and imaginary
chemical potential $\mu$ in the context of non-local chiral quark models.
One of the problems of the standard (local) PNJL model is that once the PL
potential is adjusted to reproduce the pure gauge lattice QCD results, it is
found \cite{Ratti:2005jh} that the critical temperature for the chiral and
deconfinement transitions at vanishing chemical potential, $T_c \approx
220$~MeV, is somewhat too high in comparison with the presently most
accepted lattice result, namely $T_c = 173(8)$~MeV for two light flavors
\cite{Kar03}. A solution to this difficulty follows from the observation
made in the context of the Polyakov--Quark-Meson
model~\cite{Schaefer:2007pw}, where it is claimed that in the presence of
dynamical quarks one should decrease the parameter $T_0$ which sets the
scale of the PL potential. However, in contradiction to lattice results, in
the PNJL model this sort of rescaling leads to a rather noticeable splitting
between the deconfinement and chiral restoration temperatures. This
splitting can be avoided only after the inclusion of extra eight-quark
interactions~\cite{eight}, or by assuming that the quark-quark coupling
constant is some ad-hoc function of the Polyakov loop~\cite{Sakai:2010rp}.
Here we show that in the case of the non-local SU(2) PNJL model the critical
temperature can be made naturally compatible with lattice QCD estimates,
without spoiling the entanglement between deconfinement and chiral
restoration transition temperatures even for imaginary chemical potential.
It should be stressed that the extension to imaginary chemical potential
deserves significant theoretical interest, since lattice
calculations~\cite{D'Elia:2002gd,deForcrand:2002ci,Wu:2006su} become free of
the sign problem and the corresponding results can be compared with
effective model predictions. Moreover, the behavior in the region of
imaginary chemical potential is expected to have implications on the QCD
phase diagram at finite real values of $\mu$. Lattice calculations, as well
as analyses based on the exact renormalization group
equations~\cite{Braun:2009gm}, suggest a close relation between the
deconfinement and chiral restoration transitions for imaginary chemical
potentials. Thus, we extend our study of these transitions to the region of
imaginary $\mu$, where we also analyze the characteristics of the so-called
Roberge-Weiss (RW) transition~\cite{Roberge:1986mm}, in particular, at the
RW endpoint.

\hfill

Let us briefly describe the model under consideration, namely a
non-local SU(2) chiral quark theory that includes couplings to a
background color gauge field. The Euclidean effective action is
given by~\cite{Contrera:2010kz}
\begin{equation}
S_{E}= \int d^{4}x\ \left\{
\bar{\psi}(x)\left( -i\gamma_{\mu}D_{\mu}
+\hat{m}\right)  \psi(x) -\frac{G_S}{2} \Big[ j_{a}(x)j_{a}(x)- j_{P}%
(x)j_{P}(x)\Big]+ \ {\cal U}\,(\Phi[A(x)])\right\}  \ , \label{action}%
\end{equation}
where $\psi$ is the $N_{f}=2$ fermion doublet $\psi\equiv(u,d)^T$,
$\hat{m}=m_q\,\leavevmode\hbox{\small1\kern-3.8pt\normalsize1}_{2\times2}$
is the current quark mass matrix in the isospin limit, and $D_\mu\equiv
\partial_\mu - iA_\mu$ is a covariant derivative, $A_\mu$ being color
gauge fields. The nonlocal currents $j_{a}(x),j_{P}(x)$ are given by
\begin{align}
j_{a}(x)  &  =\int d^{4}z\ {\cal G}(z)\ \bar{\psi}\left(x+\frac{z}{2}\right)
\ \Gamma_{a}\ \psi\left(  x-\frac{z}{2}\right)  \ ,\nonumber\\
j_{P}(x)  &  =\int d^{4}z\ {\cal F}(z)\ \bar{\psi}\left(  x+\frac{z}{2}\right)
\ \frac{i {\overleftrightarrow{\rlap/\partial}}}{2\ \kappa_{p}}
\ \psi\left(  x-\frac{z}{2}\right) , \label{currents}%
\end{align}
where $\Gamma_{a}=(\leavevmode\hbox{\small1\kern-3.8pt\normalsize1},i\gamma
_{5}\vec{\tau})$, and the functions ${\cal F}(z)$ and ${\cal G}(z)$ are
non-local form factors that characterize the interactions. Notice that even
if we take for convenience the same coupling parameter $G_{S}$ for both
interaction terms, the relative strength between them is controlled by the
mass parameter $\kappa_{p}$.

To proceed we perform a standard bosonization of the theory, introducing
bosonic fields $\sigma_{1,2}(x)$ and $\pi_a(x)$, and integrating out the
quark fields. We will work within the mean field approximation (MFA), in
which these bosonic fields are replaced by their vacuum expectation values
$\sigma_{1,2}$ and  $\pi_a = 0$. Since we are interested in studying the
characteristics of the chiral phase transition, we extend the bosonized
effective action to finite temperature $T$ and chemical potential $\mu$.
This can be done by using the Matsubara formalism. Concerning the gauge
fields $A_\mu$, we assume that quarks move on a constant background field
$\phi = A_4 = i A_0 = i g\,\delta_{\mu 0}\, G^\mu_a \lambda^a/2$, where
$G^\mu_a$ are the SU(3) color gauge fields. Then the traced Polyakov loop,
which is taken as order parameter of confinement, is given by
$\Phi=\frac{1}{3} {\rm Tr}\, \exp( i \phi/T)$. We will work in the so-called
Polyakov gauge, in which the matrix $\phi$ is given a diagonal
representation $\phi = \phi_3 \lambda_3 + \phi_8 \lambda_8$. This leaves
only two independent variables, $\phi_3$ and $\phi_8$. The mean field traced
Polyakov loop reads then
\begin{eqnarray}
\Phi =  \frac13 \left[ \exp\left( -\frac{2i}{\sqrt{3}}  \frac{\phi_8}{T} \right) +
                     2 \exp\left(  \frac{i}{\sqrt{3}}  \frac{\phi_8}{T}    \right)
                     \,\cos \left(\frac{\phi_3}{T} \right)
                \right]\ .
\end{eqnarray}

Within this framework the mean field thermodynamical potential $\Omega^{\rm
MFA}$ at finite temperature and arbitrary (in general, complex) chemical
potential is given by
\begin{align}
\Omega^{\rm MFA} =  \,- \,4 T \sum_{c=r,g,b} \ \sum_{n=-\infty}^{\infty} \int \frac{d^3\vec p}{(2\pi)^3}
\ \ln \left[ \frac{ (\rho_{n, \vec{p}}^c)^2 +
M^2(\rho_{n,\vec{p}}^c)}{Z^2(\rho_{n, \vec{p}}^c)}\right]+
\frac{\sigma_1^2 + \kappa_p^2\ \sigma_2^2}{2\,G_S} +
{\cal{U}}(\Phi ,\Phi^*,T) \ . \label{granp}
\end{align}
Here, $M(p)$ and $Z(p)$
are given by
\begin{eqnarray}
M(p) =  Z(p) \left[m_q + \sigma_1 \ g(p) \right] \ , \qquad
Z(p) =  \left[ 1 - \sigma_2 \ f(p) \right]^{-1}\ ,
\label{mz}
\end{eqnarray}
where $g(p)$ and $f(p)$ are Fourier transforms of ${\cal G}(z)$ and ${\cal
F}(z)$. We have also defined
\begin{equation}
\Big({\rho_{n,\vec{p}}^c} \Big)^2 =
\Big[ (2 n +1 )\pi  T- i \mu + \phi_c \Big]^2 + {\vec{p}}\ \! ^2 \ ,
\end{equation}
where the quantities $\phi_c$ are given by the relation $\phi = {\rm
diag}(\phi_r,\phi_g,\phi_b)$, i.e.\ $\phi_r = \phi_3 + \phi_8/\sqrt3$,
$\phi_g = -\phi_3 + \phi_8/\sqrt3$, $\phi_b = -2 \phi_8/\sqrt3$.

To proceed we need to specify the explicit form of the Polyakov
loop effective potential ${\cal{U}}(\Phi ,\Phi^*,T)$. Following
Ref.~\cite{Roessner:2006} we take
\begin{equation}
{\cal{U}}(\Phi ,\Phi^*,T) =
\left\{-\,\frac{1}{2}\, a(T)\,\Phi \Phi^* \;+
\;b(T)\, \ln\left[1 - 6\, \Phi \Phi^* + 4\, \Phi^3 + 4\, (\Phi^*)^3
- 3\, (\Phi \Phi^*)^2\right]\right\}\; T^4 \ ,
\end{equation}
where the coefficients are parameterized as
\begin{equation}
a(T) = a_0 +a_1 \left(\dfrac{T_0}{T}\right) + a_2\left(\dfrac{T_0}{T}\right)^2
\ ,
\qquad
b(T) = b_3\left(\dfrac{T_0}{T}\right)^3 \ .
\end{equation}
The values of $a_i$ and $b_3$ are fitted~\cite{Roessner:2006} to lattice QCD
results, which in absence of dynamical quarks lead to a deconfinement
temperature $T_0 \approx 270$ MeV. However, as mentioned above, it has been
argued~\cite{Schaefer:2007pw} that in the presence of light dynamical quarks
this value has to be modified accordingly, e.g.\ $T_0\simeq 208$ MeV for
$N_f=2$ and $T_0\simeq 180$ MeV for $N_f=3$. Effects of this change in $T_0$
will be discussed below. In addition, it is seen that $\Omega^{\rm MFA}$
turns out to be divergent, thus it has to be regularized. Here we use the
same prescription as e.g.\ in Ref.~\cite{Tum:2005}, namely
\begin{equation}
\Omega^{\rm MFA}_{\rm reg} = \Omega^{\rm MFA} - \Omega^{\rm free} + \Omega^{\rm
free}_{\rm reg} + \Omega_0 \ , \label{omegareg}
\end{equation}
where $\Omega^{\rm free}$ is obtained from Eq.~(\ref{granp}) by setting $\sigma_1
= \sigma_2=0$, and $\Omega^{\rm free}_{\rm reg}$ is the regularized expression for
the quark thermodynamical potential in the absence of fermion interactions:
\begin{equation}
\Omega^{\rm free}_{\rm reg} \ = \ -4 T \int \frac{d^3 \vec{p}}{(2\pi)^3}\;
\sum_{c=r,g,b} \ \sum_{s=\pm 1}\mbox{Re}\;
\ln\left\{ 1 + \exp\left[-\;\frac{\epsilon_p + s(\mu + i \phi_c)}{T}
\right]\right\}
\ ,
\label{freeomegareg}
\end{equation}
with $\epsilon_p = \sqrt{\vec{p}^{\;2}+m_q^2}\;$. The last term in
Eq.~(\ref{omegareg}) is just a constant fixed by the condition that
$\Omega^{\rm MFA}_{\rm reg}$ vanishes at $T=\mu=0$.

The mean field values $\sigma_{1,2}$ and $\phi_{3,8}$ can be obtained from a
set of four coupled ``gap'' equations that follow from the minimization of
the regularized thermodynamical potential,
\begin{equation}
\frac{\partial\Omega^{\rm MFA}_{\rm reg}}{\partial\sigma_{1}} \ = \
\frac{\partial\Omega^{\rm MFA}_{\rm reg}}{\partial\sigma_{2}} \ = \
\frac{\partial\Omega^{\rm MFA}_{\rm reg}}{\partial\phi_3}\ = \
\frac{\partial\Omega^{\rm MFA}_{\rm reg}}{\partial\phi_8}\ = \ 0 \ .
\label{fullgeq}
\end{equation}
Once the mean field values are obtained, the behavior of other relevant
quantities as functions of the temperature and chemical potential can be
determined. We concentrate in particular in the chiral quark condensate
$\langle\bar{q}q\rangle = \partial\Omega^{\rm MFA}_{\rm reg}/\partial
m_q$, which together with the modulus of the Polyakov loop $|\Phi|$ will
be taken as order parameters of the chiral restoration and deconfinement
transitions, respectively. For simplicity, the associated susceptibilities
will be defined as $\chi_{\rm cond} = d \langle\bar{q}q\rangle/ d T$ and
$\chi_{PL} = d |\Phi |/ d T$.

In order to fully specify the model under consideration we have to fix the
model parameters as well as the form factors $g(q)$ and $f(q)$ that
characterize the non-local interactions. Here we consider three parameter
sets A, B and C, which have been introduced in Ref.~\cite{Noguera:2008}. Set
A corresponds to the relatively simple case in which there is no wave
function renormalization (WFR) of the quark propagator, i.e.~$f(p)=0$, $Z(p)
= 1$, and $g(p)$ has a Gaussian behavior, $g(p)=\exp\left(
-p^{2}/\Lambda_{0}^{2}\right)$. In set B we consider a more general case
that includes quark WFR, taking also an exponential shape for the
corresponding form factor, $f(p)=\exp\left( -p^{2}/\Lambda_{1}^{2}\right)$.
Finally, in set C we take Lorentzian-like form factors, chosen in such a way
that one can well reproduce the momentum dependence of mass and WFR
functions obtained in lattice calculations. The parameter values for sets A,
B and C, together with the corresponding predictions for several meson
properties, can be found in Ref.~\cite{Noguera:2008}.

\hfill

Let us analyze the deconfinement and chiral transitions at vanishing
chemical potential in the framework of the model presented above. Taking
$T_0$ as a parameter, and solving numerically Eqs.~(\ref{fullgeq}) for sets
A, B and C, it is found that both the deconfinement and chiral restoration
temperatures are coincident in a wide range of values of $T_0$. This is
illustrated in Fig.~1, where we show the behavior of the relevant order
parameters and the corresponding susceptibilities for the lattice inspired
parameterization set C. We consider three characteristic values $T_0=270$,
$208$ and $180$ MeV, corresponding to the presence of 0, 2 and 3 dynamical
fermions, respectively~\cite{Schaefer:2007pw}. It is clear that for this set
both transitions are crossover-like, and they occur at basically the same
critical temperature, as it is indicated by the peaks of the corresponding
susceptibilities. One might notice that as long as $T_0$ decreases the
chiral susceptibility tends to become asymmetric around $T_c$, being
somewhat broader on the high temperature side. Though this could be
considered as an indication that for smaller values of $T_0$ chiral symmetry
is restored at a slightly higher temperature, even at $T_0=180$ MeV the
splitting between the main peak and what might be considered as a second
broad peak is less than 10 MeV. In addition, it is worth to point out that
although the coincidence of the deconfinement and chiral restoration
critical temperatures holds for all three parameter sets A, B and C, the
character of the transitions may be different from one another. This is
shown in Fig.~2, where we plot the values of the critical temperatures as
functions of $T_0$ for sets A, B and C. We see that for set A, which does
not include WFR, the transition becomes a first order one for values of
$T_0$ below $\simeq 235$ MeV. On the other hand, for the exponential
parameterization with WFR, set B, this happens at a lower value $T_0\simeq
190$ MeV. Finally, as already mentioned, for the lattice inspired
parameterization set C the transitions are crossover-like for all considered
values of $T_0$. It should be stressed that for $T_0=208$ MeV (corresponding
to our $N_f = 2$ model) the resulting critical temperatures are in good
agreement with lattice QCD estimates. Indeed, we get $T_c(0) = 173$, $171$ and
$173$ MeV for sets A, B and C, respectively. It is also important to remark
that the nature of deconfinement and chiral restoration transitions for two
light flavors in lattice QCD is still under debate. While most
studies~\cite{Karsch:1994hm,Bernard:1996iz,Iwasaki:1996ya,Aoki:1998wg,Ali
Khan:2000iz} favor a second order transition in the chiral limit, there are
also claims for a first order transition~\cite{D'Elia:2005bv,Bonati:2009yg}.
Given that in the context of the present non-local model the
parameterizations that include WFR appear to be more realistic, the second
order scenario turns out to be preferred.

We consider now the situation at nonzero imaginary chemical potential. As it
is well known, Roberge and Weiss found~\cite{Roberge:1986mm} that the
thermodynamical potential of QCD in presence of an imaginary chemical
potential $\mu = i\, \theta\, T$ is a periodic function of $\theta$ with
period $2 \pi/3$. This means that QCD is invariant under a combination of a
$Z_3$ transformation of the quark and gauge fields and a shift $\theta
\rightarrow \theta + 2\, k\, \pi/3$, with integer $k$, in the chemical
potential. Recently, it has been shown that this so-called extended $Z_3$
transformation is also a symmetry of the local Polyakov-Nambu-Jona-Lasinio
model~\cite{Sakai:2008um}. Indeed, in the context of this model the
thermodynamical potential is invariant under the transformations
\begin{eqnarray}
\Phi(\theta) \rightarrow  \Phi(\theta) \exp(-i\, 2\,k\, \pi/3) \ , \qquad
\Phi^*(\theta) \rightarrow  \Phi^*(\theta) \exp(i\, 2\, k\, \pi/3)  \ , \qquad
\theta \rightarrow \theta + 2\, k\, \pi/3\ .
\label{extz3}
\end{eqnarray}
The RW periodicity is a remnant of the $Z_3$ symmetry in the pure gauge
limit. In QCD with dynamical quarks, if the temperature is larger than a
certain value $T_{RW}$ it can be seen that three $Z_3$ vacua appear. These
vacua can be classified through their Polyakov loop phases, given by
$\varphi$, $\varphi+2 \pi/3$ and $\varphi+4 \pi/3$. Roberge and Weiss showed
that for $T > T_{RW}$ there is a first order phase transition at
$\theta=\pi/3$ mod $2 \pi/3$, in which the vacuum jumps to one of its $Z_3$
images. This is known as the ``Roberge-Weiss transition'', and the point at
the end of the RW transition line in the $(T,\theta)$ plane,
i.e.~$(T,\theta) = (T_{RW}, 2 \pi/3)$, is known as the ``RW endpoint''. The
order of the RW transition at the RW endpoint has been subject of
considerable interest recently in the framework of lattice
QCD~\cite{D'Elia:2009qz,deForcrand:2010he,Bonati:2010gi} due to the
implications it might have on the QCD phase diagram a finite real $\mu$.
According to lattice calculations, it appears that for two light flavors the
RW endpoint is first order for realistically small values of the current
quark mass. Following these considerations it is important to check whether
the thermodynamical potential of the non-local PNJL models studied in this
work does respect the extended $Z_3$ symmetry. In fact, it is easy to show
that this is the case. The last two terms in Eq.~(\ref{granp}) are obviously
invariant under the transformations in Eq.~(\ref{extz3}), whereas to check
the invariance of the first term it is convenient to write these
transformations in the equivalent way
\begin{eqnarray}
\phi_3(\theta) \rightarrow  \phi_3(\theta) \ , \qquad \phi_8(\theta)
\rightarrow  \phi_8(\theta) - 2 \, k\,\pi\,T/\sqrt{3} \ , \qquad \theta
\rightarrow \theta + 2\, k\, \pi\,/3\ .
\label{extz3bis}
\end{eqnarray}
Thus it can be easily proven that any sum of the form $\sum_{c=r,g,b}
F[(\rho_{n, \vec{p}}^c)^2]$, where $F$ is an arbitrary function, turns out
to be invariant under the extended $Z_3$ transformations. The invariance of
the terms introduced in the regularization procedure [c.f.
Eq.(\ref{omegareg})] can be shown in the same way.

\begin{figure}[hbt]
\includegraphics[width=0.55 \textwidth]{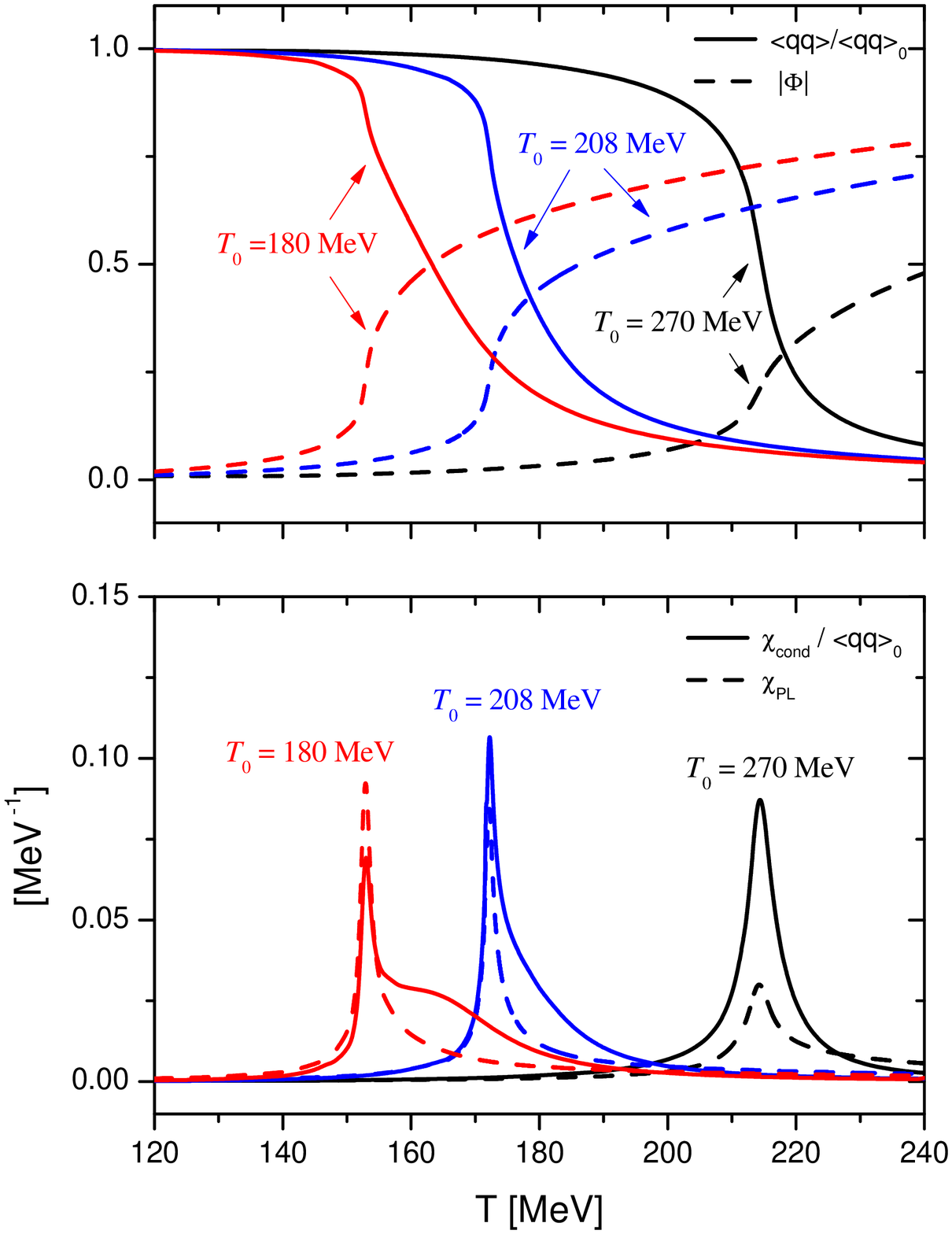}
\caption{Order parameters for the deconfinement and chiral restoration
transitions (upper panel) and corresponding susceptibilities (lower panel)
as functions of $T$ for three characteristic values of $T_0$, namely $T_0
= 270$, 208 and 180 MeV. Curves correspond to Set C.} \label{fig1}
\end{figure}

Having checked that our non-local PNJL models possess the extended $Z_3$
invariance we turn now to the results of the numerical analysis of the
behavior of the different order parameters as functions of $T$ and $\theta$
using the value $T_0=208$ MeV corresponding to two light flavors. We first
keep $T$ fixed, verifying that the expected periodicity of the different
thermodynamical quantities as functions of $\theta$ is indeed satisfied.
Moreover, for $T > T_{RW}$ we find the mentioned RW first order phase
transition at $\theta=\pi/3$, which is signalled by a discontinuity in the
phase of the Polyakov loop field. The values obtained for $T_{RW}$ are $191$
MeV, $188$ MeV and $191$ MeV for sets A, B and C, respectively, in good
agreement with the lattice QCD estimate $T_{RW}=185(9)$~\cite{Wu:2006su}.
Concentrating on the sector $0 \le \theta \le \pi/3$ we observe that for
values of the temperature $T_c(\theta = 0) \le T \le T_{RW}$ the order
parameters for both deconfinement and chiral symmetry show signals of a
phase transition at a given value of $\theta$. This is clearly seen in
Fig.~\ref{fig3} where we plot the behavior of order parameters and
susceptibilities as functions of $T$ taking now $\theta$ fixed at two
representative values $\theta=\pi/6$ and $\pi/3$. The plots correspond to
parameter set C. We note that while for $\theta = \pi/6$ both deconfinement
and chiral restoration are crossover-like, they are first order for $\theta
=\pi/3$. As in the case of $\theta=0$ (see the curves corresponding to
$T_0=208$~MeV in Fig.~\ref{fig1}), for both values of $\theta$ the
deconfinement and chiral restoration transitions occur at the same
temperature, given by the peaks of the susceptibilities or the positions of
the discontinuities. Although one might argue that there is a certain
tendency of the chiral susceptibility to decay more slowly or even, in the
case of $\theta=\pi/3$, to display a very broad peak on the high temperature
side, one can hardly conclude that both transitions apart from each other
more than about $20$ MeV, even for $\theta = \pi/3$. The situation is quite
similar for parameter set B (which also includes WFR), whereas for set A one
gets just first order transitions for all values of $\theta$ in the range of
temperatures considered. The dependence of the critical temperature $T_c$ as
a function of $\theta$ is shown in Fig.~\ref{fig4} for our three parameter
sets. For comparison, we also show the corresponding lattice results given
in Ref.~\cite{Wu:2006su}, which include an error of about $10\%$ due to the
uncertainty in the lattice determination of $T_c(\theta = 0)$. As already
mentioned, while for set A both the deconfinement and chiral restoration
transitions are always first order, in the case of sets B and C there is a
critical value $\theta \sim 0.7 \times \pi/3$ below which the transitions
become crossover-like. Thus, we find that for all three parameterizations
the corresponding transition lines are first order when they meet the RW
endpoint. This implies that the RW endpoint is a triple point, the RW
transition being also first order there. The character of the RW transition
at the RW endpoint is clearly seen in Fig.~\ref{fig5}, where we plot the
behavior of the phase of the extended Polyakov loop $\Psi$ at $\theta=\pi/3$
as a function of $T$ (the figure corresponds to set C). The extended
Polyakov loop, defined by $\Psi = \exp(i\theta)\Phi$, is by construction
invariant under the transformations in Eq.~(\ref{extz3}), and its phase
$\psi$ can be taken as order parameter of the RW
transition~\cite{Sakai:2008um}.

In summary, we have considered the impact of the feedback of the dynamical
quarks on the parameter $T_0$ appearing in the Polyakov potential, as
proposed in Refs.~\cite{Schaefer:2007pw}. This has been done here in the
context of non-local PNJL models, considering three different types of
non-local form factors. We have studied the deconfinement and chiral
restoration transitions, determining the corresponding critical
temperatures and the character of the transitions. The results are found
to be in agreement with those obtained in lattice QCD, showing a natural
entanglement between both critical temperatures for zero and imaginary
chemical potential. We have also analyzed the Roberge Weiss transition,
which is found to be first order at the RW endpoint for all three
parameter sets.

This work has been partially supported by CONICET (Argentina) under grants
\# PIP 00682 and PIP 02495, and by ANPCyT (Argentina) under grant \# PICT07
03-00818.

\pagebreak

\begin{figure}[hbt]
\includegraphics[width=0.4 \textwidth]{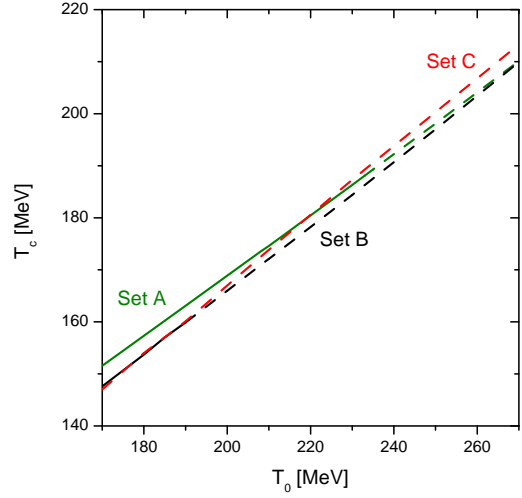}
\caption{Critical temperature as a function of $T_0$ for our parameter
sets A, B and C. Solid and dashed lines stand for first order and
crossover-like transitions, respectively. Note that below $T_0 \sim 190$
MeV the curves for sets B and C almost overlap. However, while for set B
the transition in that region  is first order (solid line), for set C it
is crossover-like (dashed line).} \label{fig2}
\end{figure}

\begin{figure}[hbt]
\includegraphics[width=0.5 \textwidth]{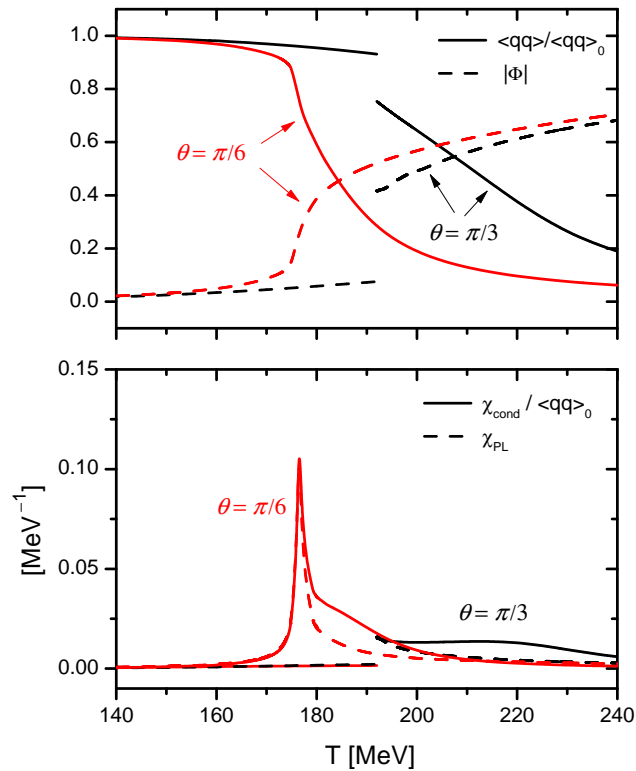}
\caption{Order parameters for the deconfinement and chiral restoration
transitions (upper panel) and corresponding susceptibilities (lower panel)
as functions of $T$ for $\theta=\pi/6$ and $\theta=\pi/3$. Curves
correspond to Set C.} \label{fig3}
\end{figure}

\begin{figure}[hbt]
\includegraphics[width=1.0 \textwidth]{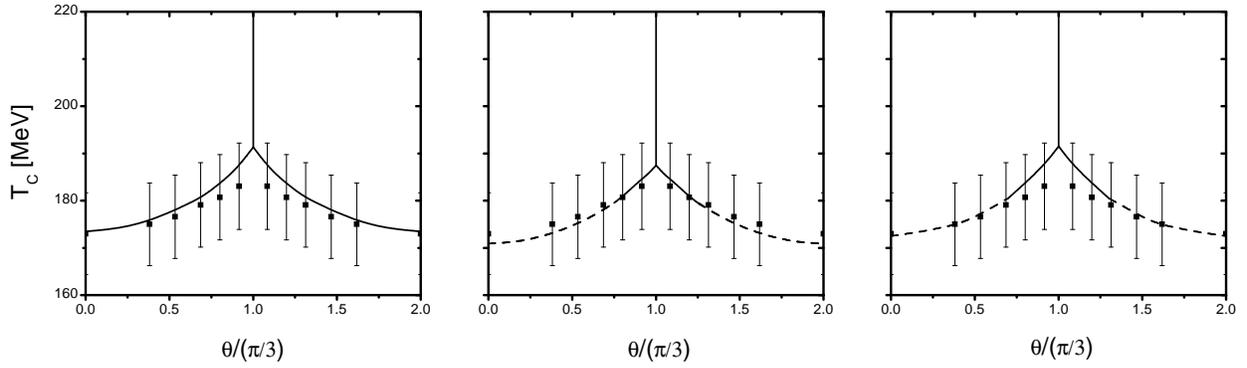}
\caption{Critical temperature as a function of $\theta$ for
parameter sets A (left), B (center) and C (right). Solid and
dashed lines stand for first order and crossover-like transitions,
respectively. Dots correspond to lattice QCD results. Vertical
solid lines correspond to the first order RW transition.}
\label{fig4}
\end{figure}

\begin{figure}[hbt]
\includegraphics[width=0.6 \textwidth]{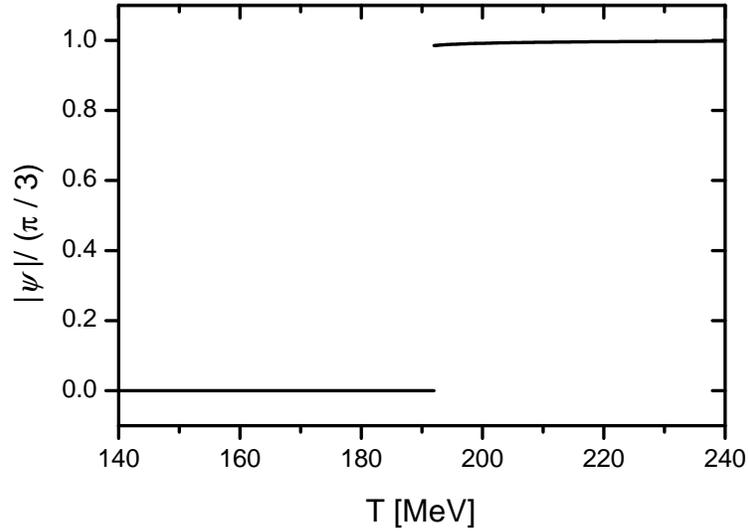}
\caption{Phase of the extended Polyakov loop $\Psi = \exp(i \theta)\Phi$
as a function of $T$ for $\theta=\pi/3$. Curves correspond to set C.}
\label{fig5}
\end{figure}

\end{document}